\documentclass[aps,prb,twocolumn,superscriptaddress,showpacs,superbib,floats]{revtex4}
\usepackage{graphicx}
\usepackage{amssymb}
\usepackage{amsmath}
\usepackage{dcolumn}
\usepackage{color}

\begin{document}

\title{
Characterization of the Spin-1/2 Linear-Chain Ferromagnet CuAs$_2$O$_4$}

\author{K. Caslin}
\email[Corresponding author:~E-mail~]{k.caslin@fkf.mpg.de}
\affiliation{Max-Planck-Institut f{\"u}r Festk{\"o}rperforschung, Heisenbergstrasse 1, D-70569 Stuttgart, Germany}
\affiliation{Brock University, 500 Glenridge Ave., St. Catharines, Ontario L2S-3A1, Canada}
\author{R. K. Kremer}
\affiliation{Max-Planck-Institut f{\"u}r Festk{\"o}rperforschung, Heisenbergstrasse 1, D-70569 Stuttgart, Germany}
\author{F. S. Razavi}
\affiliation{Brock University, 500 Glenridge Ave., St. Catharines, Ontario L2S-3A1, Canada}
\author{A. Schulz}
\affiliation{Max-Planck-Institut f{\"u}r Festk{\"o}rperforschung, Heisenbergstrasse 1, D-70569 Stuttgart, Germany}
\author{A. Mu\~{n}oz}
\affiliation{MALTA Consolider Team, Departamento de F\'{\i}sica Fundamental II, and Instituto de Materiales y Nanotecnolog\'{\i}a, Universidad de La Laguna, La Laguna 38205, Tenerife, Spain}

\author{F. Pertlik}
\affiliation{Vienna University of Technology, Institute of Mineralogy and Crystallography, Althanstr. 14, A-1090 Wien, Austria}

\author{J. Liu}
\author{M.-H. Whangbo}
\affiliation{Department of Chemistry, North Carolina State
University, Raleigh, North Carolina 27695-8204, U.S.A.}

\author{J. M. Law}
\affiliation{Hochfeld-Magnetlabor Dresden, Helmholtz-Zentrum Dresden-Rossendorf, D-01314 Dresden, Germany}

\date{\today}

\begin{abstract}
The magnetic and lattice properties of the $S$=1/2 quantum-spin-chain ferromagnet, CuAs$_2$O$_4$, mineral name trippkeite, were investigated.
The crystal structure of CuAs$_2$O$_4$ is characterized by the presence of corrugated CuO$_2$ ribbon chains.
Measurements of the magnetic susceptibility, heat capacity, electron paramagnetic resonance and Raman spectroscopy were performed.
Our experiments conclusively show that a ferromagnetic transition occurs at $\sim$7.4 K.
\textit{Ab initio}  DFT calculations reveal dominant ferromagnetic nearest-neighbor and weaker antiferromagnetic next-nearest-neighbor spin exchange interactions along the ribbon chains. The ratio of $J_{\rm nn}$/$J_{\rm nnn}$ is near -4, placing  CuAs$_2$O$_4$ in close proximity to a quantum critical point in the $J_{\rm nn}$ - $J_{\rm nnn}$ phase diagram.
TMRG simulations used to analyze the magnetic susceptibility confirm this ratio.
Single-crystal magnetization measurements indicate that a magnetic anisotropy forces the Cu$^{2+}$ spins to lie in an easy plane perpendicular to the $c$-axis. An analysis of the field and temperature dependent magnetization by modified Arrott plots  reveals a 3d-XY critical behavior.
Lattice perturbations induced by quasi-hydrostatic pressure and temperature were mapped via magnetization and Raman spectroscopy.

\end{abstract}

 \maketitle


\email{k.caslin@fkf.mpg.de}

\section{Introduction}\label{SecIntro}

Low dimensional magnetic Cu$^{2+}$ systems containing CuX$_2$ ribbon chains
have attracted a great deal of attention because of their unusual intrachain nearest- and next-nearest neighbor spin exchange relations.
In such compounds it is frequently  found that the spin exchange interactions between the Cu$^{2+}$ ($S$=1/2) ions are such that the next-nearest-neighbor (NNN) exchange is antiferromagnetic (AFM), the nearest-neighbor (NN) exchange is ferromagnetic (FM), and the NNN spin exchange is often considerably stronger than the NN spin exchange. Due to the inherent competition  of the NN and NNN spin exchange interactions, the CuX$_2$ ribbon chains tend to develop unusual AFM incommensurate spiral spin structures\cite{Gibson2004,Enderle2005,Capogna2005,Drechsler2007a,Banks2009,Law2011,Lee2012,Willenberg2012,Wolter2012}
and sometimes concomitantly multiferroic behavior.\cite{Naito2007,Schrettle2008,Yasui2008,Zhao2012}
These CuX$_2$ ribbon chains are formed by linking  CuX$_4$ basal-square-planes of axially elongated CuX$_6$ (X = O, Cl, Br,...) octahedra together via their trans-edges.

The magnetic properties of the CuX$_2$ ribbon chains are primarily determined by the ratio of the NN to NNN spin exchange parameters, $J_{\rm nn}$ and $J_{\rm nnn}$, with $\alpha$ = $J_{\rm nn}$/$J_{\rm nnn}$. However, at low temperatures additional interactions (e.g., smaller interchain spin exchange interactions) also become important. Interchain interactions usually drive the systems to long-range magnetic order, the exact details of which are often determined by additional weak magnetic anisotropies.\cite{Enderle2005,Mourigal2011,Mourigal2012}
Over the last decade, much interest has been devoted to such ribbon chain systems with spin exchange parameters lying within the so-called frustrated regime, i.e., between the Majumdar-Ghosh point, $\alpha$ = 2, and the 'FM point', $\alpha$ = -4. A first-order phase transition to a FM ground is expected at $\alpha$ = -4 and hence, it constitutes a quantum critical point (QCP) in the vicinity of which small perturbations, such as interchain exchange and anisotropic exchange couplings, may induce a pronounced response of the system.\cite{Bursill1995,White1996,Sachdev1999}

Several CuO$_2$ ribbon chain systems with $\alpha\sim$-4 have been investigated, none of which undergoes long-range FM ordering at low temperatures.
Some  systems, such as Li$_2$ZrCuO$_4$ and PbCuSO$_4$(OH)$_2$ (linarite), prevent long-range FM ordering since they exhibit an AFM incommensurate spin-spiral structure along the ribbon chains.\cite{Drechsler2007b,Schmitt2009,Willenberg2012,Wolter2012}
Other systems, such as Ca$_2$Y$_2$Cu$_5$O$_{10}$ and Li$_2$CuO$_2$, do contain FM ribbon chains, however, weak interchain interactions force the FM-ribbon chains to align antiparallel, resulting in long-range AFM ordering.\cite{Matsuda2001,Kuzian2012,Sapina1990,Boehm1998, Xiang2007,Drechsler2010}  In the case of Li$_2$CuO$_2$, the absence of a spiral magnetic order in the ribbon chains is understood to be a consequence of order by disorder arising from interchain interactions.\cite{Xiang2007} In all these cases, it appears that either magnetic spiral ordering or weak interchain interactions drive the systems to long-range AFM ordering.
It also appears to be possible for such systems not to exhibit a magnetic long-range ordering at all. For example, the system Rb$_2$Cu$_2$Mo$_3$O$_{12}$, with $\alpha \sim$ -2.7, does not undergo long-range magnetic ordering down to 2 K even though magnetic field induced ferroelectricity is observed below 8 K.\cite{Hase2004,Yasui2013}

\begin{figure}[htp]
\includegraphics[width=6cm]{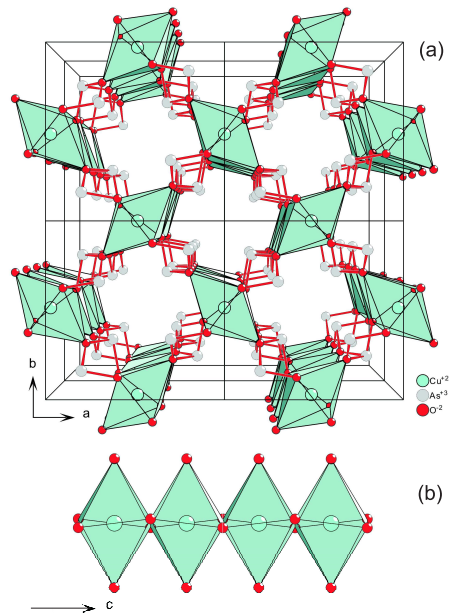}
\caption{(color online) (a) Projection along the [001] direction of the trippkeite crystal structure . The Cu$^{2+}$ atoms are represented by the large (green) spheres, the oxygen atoms by small (red) spheres, and the As atoms by (grey) medium spheres. (b) A corrugated chain of trans-edge connected CuO$_6$ octahedra highlighting the twisted basal planes of the octahedra in CuAs$_2$O$_4$, which lead to a corrugation of the CuO$_2$ ribbon chains.}
\label{Fig1}
\end{figure}

Here we describe the magnetic and lattice properties of a new system, trippkeite, featuring edge-sharing CuO$_2$ ribbon chains. Trippkeite is an exceptional ribbon chain system because it shows a FM ground state below $\sim$7.4 K.
Trippkeite is a natural  mineral with the chemical composition CuAs$_2$O$_4$.
A first investigation and description of the crystal structure of  natural trippkeite, by Zemann in 1951, was based on the assumption that trippkeite is isotypic to schafarzikite, FeSb$_2$O$_4$.\cite{Zemann1951}  In 1975, Pertlik employed hydrothermal synthesis to prepare small  crystals of synthetic trippkeite and performed a full crystal structure determination which confirmed Zemann's earlier results (see Figure \ref{Fig1}).\cite{Pertlik1975,Pertlik1977} This investigation showed that trippkeite is isostructural to the family of compounds with the general composition MT$_2$O$_4$ (M$^{2+}$ = Mg, Mn, Fe, Co, Ni, Zn; T$^{3+}$ = As, Sb, Bi), with schafarzikite, FeSb$_2$O$_4$, being the most popular member.\cite{Kumada2009}

Trippkeite crystallizes in a tetragonal structure (space group $P$4$_2$/$mbc$) with lattice parameters $a$ = $ b$ = 8.592(4) $\rm \AA$  and $c$ = 5.573(4) $\rm \AA$.\cite{Pertlik1975}
The Cu atoms are located on Wyckoff sites 4$d$ and the As atoms on sites 8$h$. The O atoms occupy Wyckoff sites 8$g$(O1) and 8$h$(O2). The Cu atoms are coordinated by elongated oxygen octahedra (Cu - O distances: 4 $\times$ 1.95 \AA; 2 $\times$ 2.47 \AA), which connect via trans-edges to form corrugated CuO$_2$ ribbons. The O1 and O2 atoms are located at the apical and basal positions of the octahedra, respectively. The As atoms form AsO$_3$ pyramids, which link the oxygen atoms in the basal planes with the apical oxygen atoms of neighboring chains such that the basal planes of neighboring chains are perpendicular to each other.\cite{Pertlik1975}  The $4s^2$ electrons of the As atoms act as pseudo-ligands and extend into the channels enclosed by four neighboring chains. Similar spacious structures are known to have a high susceptibility to structural phase transitions that can be induced by external pressure.\cite{Hinrichsen2006}

\section{Experimental Details}\label{SecExperimental}
Small wine-bottle-green crystals of CuAs$_2$O$_4$, approximately 1 $\times$ $10^{-3}$ mm$^3$ in size, were prepared by hydrothermal synthesis as described in detail by Pertlik.\cite{Pertlik1977} Some of the crystals showed brownish impurity appendages, such crystals were discarded.
Sample purity was checked by X-ray powder diffraction on crushed crystals. A STOE STADI-P diffractometer with Mo $K\alpha _1$ radiation, $\lambda$ = 0.7093 {\AA}, was used.

Temperature and magnetic field dependent magnetizations were measured with a Quantum Design, superconducting quantum interference device magnetometer (MPMS). A selection of randomly oriented crystals (mass $\sim$ 5.3 mg) were filled in a quartz glass tube, fixed with a minute amount of a fast drying varnish and measured between 1.85 K and 375 K. Directional dependent magnetizations versus field at constant temperatures were collected on a well defined large single crystal (mass 8 $\pm 2 \mu$g), oriented by x-ray diffraction with the $c$-axis parallel and perpendicular to the magnetic field.

Magnetization measurements under pressure were performed in
a copper-beryllium pressure clamp cell providing hydrostatic pressure up to 1.2 GPa and using silicon oil as a pressure medium. The pressure was determined from the superconducting critical temperature of a high purity (99.999\%) Sn sample located next to the CuAs$_2$O$_4$ sample within the pressure cell.\cite{Eiling1981}

Heat capacity measurements were performed in a Quantum Design, physical property measurement system calorimeter (PPMS). A collection of randomly oriented crystals (mass $\sim$ 3.4 mg) were thermally anchored to the calorimeter platform using a minute amount of Apiezon N grease. The heat capacity of the sample platform with grease was determined in a preceding run and subtracted from the total heat capacities.

Raman spectra were measured with a Jobin Yvon Typ V 010 LabRAM single grating spectrometer with $\sim$1 cm$^{-1}$ spectral resolution. The spectrometer setup was equipped with a double super razor edge filter, Peltier cooled CCD camera and a Mikrocryo cryostat with a copper cold finger. Measurements were performed with linearly polarized He/Ne gas laser light of 632.817 nm with $<$ 1mW of power. The light beam was focused to a 10 $\mu$m spot on the top surface of the sample using a microscope. Orientated measurements were taken parallel and perpendicular to the  $c$-axis in temperatures ranging between 4 K and 325 K.

Electron paramagnetic resonance (EPR) measurements were carried out with a Bruker ER040XK X-band microwave spectrometer and Bruker BE25 magnet controlled by a BH15 field controller calibrated with Diphenylpicrylhydrazyl (DPPH). The spectra of a selection of non-oriented crystals (mass $\sim$ 4.5 mg) were collected  with microwave radiation of $\sim$9.48 GHz in  temperature cycles ranging from  15 K to 275 K.

\section{Theoretical Details}\label{SecTheory}
\subsection{Spin Exchange Interactions}
\begin{figure}[htp]
\includegraphics[width=8cm ]{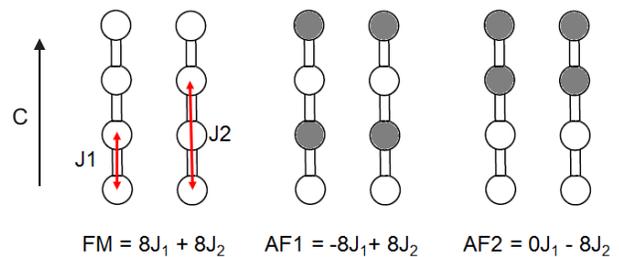}
\caption{(color online) Three order spin states of CuAs$_2$O$_4$ used to determine the values of $J_{\rm nn}$ ($\equiv J_1$) and $J_{\rm nnn}$ ($\equiv J_2$)  by DFT+$\textit{U}$ calculations. Only the Cu sites are shown for simplicity. The unfilled and filled circles represent up-spin and down-spin Cu$^{2+}$ sites, respectively.}
\label{Fig2}
\end{figure}

The intrachain spin exchange interactions, $J_{\rm nn}$ and $J_{\rm nnn}$, of CuAs$_2$O$_4$ were evaluated by performing energy-mapping analyses \cite{map1,map2} based on first principles DFT calculations for the three ordered spin states depicted in Figure \ref{Fig2}.
The energies of the three order states can be written in terms of the Heisenberg spin Hamiltonian;
\begin{equation}
\cal{H}=-\rm \sum{\textit{J}_{ij}\vec{\textit{S}_i}\vec{\textit{S}_j}},
\label{HeisHam}
\end{equation}
where \textit{J}$_{ij}$ are the exchange parameters for the coupling between spin sites \textit{i} and \textit{j}. According to the energy expressions for spin dimers with $N$ (= 1 in this case) unpaired spins per spin site \cite{Dai2001,Dai2003}, the total spin exchange energies of the three ordered spin states, per eight formula units (FUs), are expressed as summarized in Figure \ref{Fig2}.
We calculated the electronic energies of the three ordered spin states by employing the projected augmented-wave (PAW) method\cite{Blochl1994,Kresse1999} encoded in the Vienna \textit{ab~initio} simulation package (VASP) \cite{Kresse1996} and the generalized gradient approximation (GGA) for the exchange and correlation functional. \cite{Perdew2008} The plane-wave cut-off energy was set to 400 eV and a set of 18 $k$-points for the irreducible Brillouin zone was used.
To probe the effect of electron correlations associated with the Cu 3$d$ state, we performed DFT plus on-site repulsion (DFT+$\textit{U}$) calculations with $U_{\rm eff}$ = 0, 4, 6 and 8 eV for Cu.\cite{Dudarev1998} By mapping the relative energies of the three ordered spin states obtained from our DFT+$U$ calculations onto the corresponding Heisenberg Hamiltonian (Eq. (\ref{HeisHam})), we obtain the values of the nearest- and next-nearest neighbor spin exchange parameters, $J_{\rm nn}$ ($\equiv J_1$) and $J_{\rm nnn}$ ($\equiv J_2$).
\begin{table}
     \caption{Values of the NN and NNN spin exchange constants, $J_{\rm nn}$ and $J_{\rm nnn}$, respectively, obtained from the DFT+$\textit{U}$ calculations along with the Curie-Weiss temperatures calculated using Eq. (\ref{Eq2}).}
   \centering
  \begin{tabular}{ c  c  c  c  }
\hline
\hline
    $U_{\rm eff}$ (eV) & $J_{\rm nn}$ (K) & $J_{\rm nnn}$ (K) &  $\Theta_{\rm CW}$ (K) \\
\hline
    0 & 42.3 & -25.9 & 8.2 \\
    4 & 38.8 & -13.5 & 12.7 \\
    6 & 34.0 & -10.0 & 12.0 \\
    8 & 27.5 & -7.1 & 10.2 \\
\hline
\hline
  \end{tabular}
  \label{Table1}
\end{table}
\begin{figure}[htp]
\includegraphics[width=8cm ]{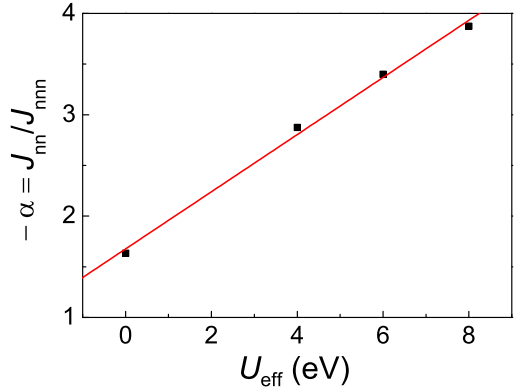}
\caption{(color online) The ratio of the NN to NNN spin exchange constants of CuAs$_2$O$_4$ calculated from the DFT+\textit{U} calculations as a function of $U_{\rm eff}$. The plot displays the dominance of the FM $J_{\rm nn}$ term over the AFM $J_{\rm nnn}$ term.}
\label{Fig3}
\end{figure}

The results summarized in Table \ref{Table1} show that the FM-NN spin exchange dominates over the AFM-NNN spin exchange.
As shown in Figure \ref{Fig3}, the ratio of the NN over the NNN spin exchanges increases with increasing $U_{\rm eff}$ values used in the DFT+$U$ calculations.
In general, the spin exchange $J$ between two spin sites (say, 1 and 2 represented by the magnetic orbitals  $\Phi_1$ and  $\Phi_2$, respectively) is written as $J$ = $J_{\rm F}$ + $J_{\rm AF}$. The FM component $J_{\rm F}$ increases when increasing the overlap density   $\Phi_1\,\Phi_2$ while the AFM component $J_{\rm AF}$ is proportional to $t^2$ and inversely proportional to $U_{\rm eff}$.\cite{map1,map2}
The magnetic orbital, $x^2$ - $y^2$ orbital, of the Cu$^{2+}$ ion has large O 2$p$ contributions.
The strong FM-NN interaction in CuAs$_2$O$_4$ is traced to the fact that the bond angle of the Cu-O-Cu superexchange path is close to 90$^\circ$ ($\sim$ 91.5$^\circ$) and the magnetic orbitals of the NN Cu$^{2+}$ ions lead to a large overlap density around the bridging oxygen atoms.\cite{map2} The weaker AFM-NNN interaction is a consequence of the twisting of the CuO$_2$ ribbon chains since it reduces the hopping integral between the NNN Cu$^{2+}$ ions.

The Curie Weiss temperatures listed in Table \ref{Table1} were calculated with the equation,
\begin{equation}
\Theta_{\rm CW}=\frac{1}{3}S(S+1)\sum_i z_iJ_i.
\label{Eq2}
\end{equation}
The $J_i$'s represent the NN and the NNN spin exchange interactions along the ribbon chains, $J_{\rm nn}$ and $J_{\rm nnn}$, respectively. $z_i$ is the number of neighbors with spin exchange $J_i$ in the NN and NNN shell, $z_{\rm nn}$ =  $z_{\rm nnn}$ = 2 for CuAs$_2$O$_4$. $\Theta_{\rm CW}$ is positive if the spin exchange is predominantly FM. The calculated Curie-Weiss temperatures are positive, consistent with the experimental findings (see below).
The $U_{\rm eff}$ values of 6 and 8 eV, most appropriate for Cu$^{2+}$, indicate -3.9 $< \alpha <$ -3.4 (see Fig. 3), close to the FM-QCP at $\alpha$ = -4. Since the $x^2$ - $y^2$ magnetic orbitals of neighboring ribbon chains are largely orthogonal to each other,   the interchain spin exchange interactions  are expected to be small and not easily accessible with DFT calculations.

\subsection{TMRG Calculations}
\begin{figure}[ht]
	\centering
		\includegraphics[width=8cm]{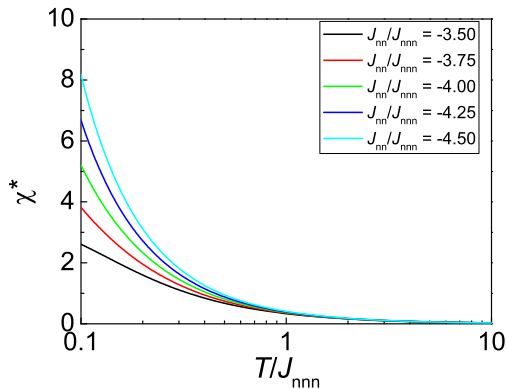}
		\caption{(Color online) TMRG spin susceptibilities, $\chi^*$, versus temperature for various ratios of the NN to NNN spin exchange interactions as indicated in the inset. $\chi^*$ = $\chi_{\rm mol} J_{\rm nnn}$/$N_{\rm A} \mu_{\rm B}^2 g^2$.}
\label{Fig4a}
\end{figure}

The temperature dependent magnetic susceptibilities, $\chi^*$ (see Figure \ref{Fig4a}), of a NN--NNN spin exchange, Heisenberg $S$=1/2 spin-chain was  simulated via transfer-matrix
density-matrix renormalization group (TMRG) calculations,\cite{Bursill1996} as implemented by Wang $et~al$.\cite{Wang1997} A consistent set of parameters was used for the simulations of $\chi^{\rm{\ast}}$ in the reduced temperature range 0.1$\leq$ ${T}/{J_{\rm{nnn}}}$ $\leq$10 and the $\alpha$ range  -4.50 $\leq {J_{\rm{nn}}}/{J_{\rm{nnn}}}\leq$ -3.50. 150 states were kept with ${H}/{J_{\rm{nnn}}}$=0.001,  $J_{\rm{nnn}}$ = 1 and a maximum trotter number of 4000. No difference was seen between 250 and 150 states down to ${T}/{J_{\rm{nnn}}}$=0.1 (not shown here), as such, 150 states were retained.

\subsection{Lattice Properties}
The lattice properties were obtained from DFT calculations using the VASP code combined with the phonon package.\cite{Parlinski}
The PAW scheme within the VASP package accommodates the full nodal character of an all-electron charge density in the core region. To achieve highly converged results with an accurate description of the electronic and dynamical properties, basis sets with plane waves up to a 520 eV cut-off energy were used. A GGA with PBEsol prescription was utilized to describe the exchange-correlation energy.\cite{Perdew2008}  In order to obtain highly converged energies and forces, a dense special $k$-point sampling for the integration of the Brillouin zone was performed.
At each selected volume, the structures were fully relaxed to their equilibrium configurations through the calculation of the forces on the atoms and stress tensor.\cite{Mujica2003} Lattice-dynamic calculations of phonon modes were performed at the zone center ($\Gamma$-point) of the Brillouin zone using the direct force-constant approach (or the supercell method).\cite{Parlinski1997}  These calculations provide information about the symmetry of the modes and their polarization vectors, and also allowed us to identify the irreducible representations and the character of the phonon modes at the $\Gamma$-point.
The calculated Raman frequencies with their  assigned symmetries are listed in Table \ref{Table3} in comparison to the experimental observations.

\begin{figure}[htp]
\includegraphics[width=8cm ]{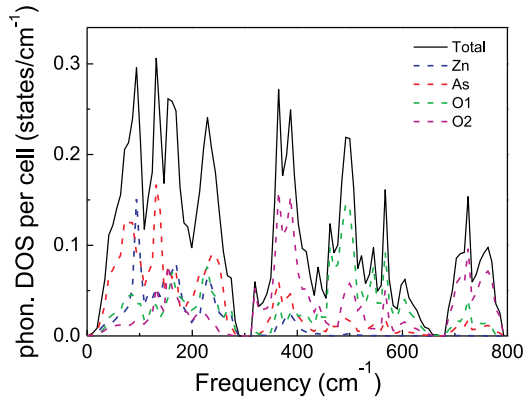}
\caption{(color online) Total and partial phonon densities of states of a hypothetical compound ZnAs$_2$O$_4$ per unit cell.}
\label{Fig4}
\end{figure}

To gain more insight on the lattice properties  of CuAs$_2$O$_4$, DFT-GGA and LDA calculations were performed for a hypothetical, diamagnetic compound ZnAs$_2$O$_4$, discarding any magnetic contributions. Figure \ref{Fig4} displays the total and partial phonon densities of states of the hypothetical compound ZnAs$_2$O$_4$. The phonon spectrum is characterized by a set of rather sharp bands  indicating nearly localized lattice vibrations extending up to $\sim$800 cm$^{-1}$. The set of phonon bands decomposes  into three subgroups, one group below $\sim$300 cm$^{-1}$, corresponding mainly to Zn and As vibrations with little contribution from O vibrations. Above $\sim$300  cm$^{-1}$ two groups of phonon bands corresponding mainly to O vibrations with almost no contribution from the heavier Zn and As atoms. Vibrations related to O atoms in the basal plane of the distorted octahedra (O2) exhibit the  highest vibrational frequencies.

\section{Results and Discussion}\label{Results}

\subsection{Raman Scattering}
\begin{figure}[htp]
\includegraphics[width=8cm ]{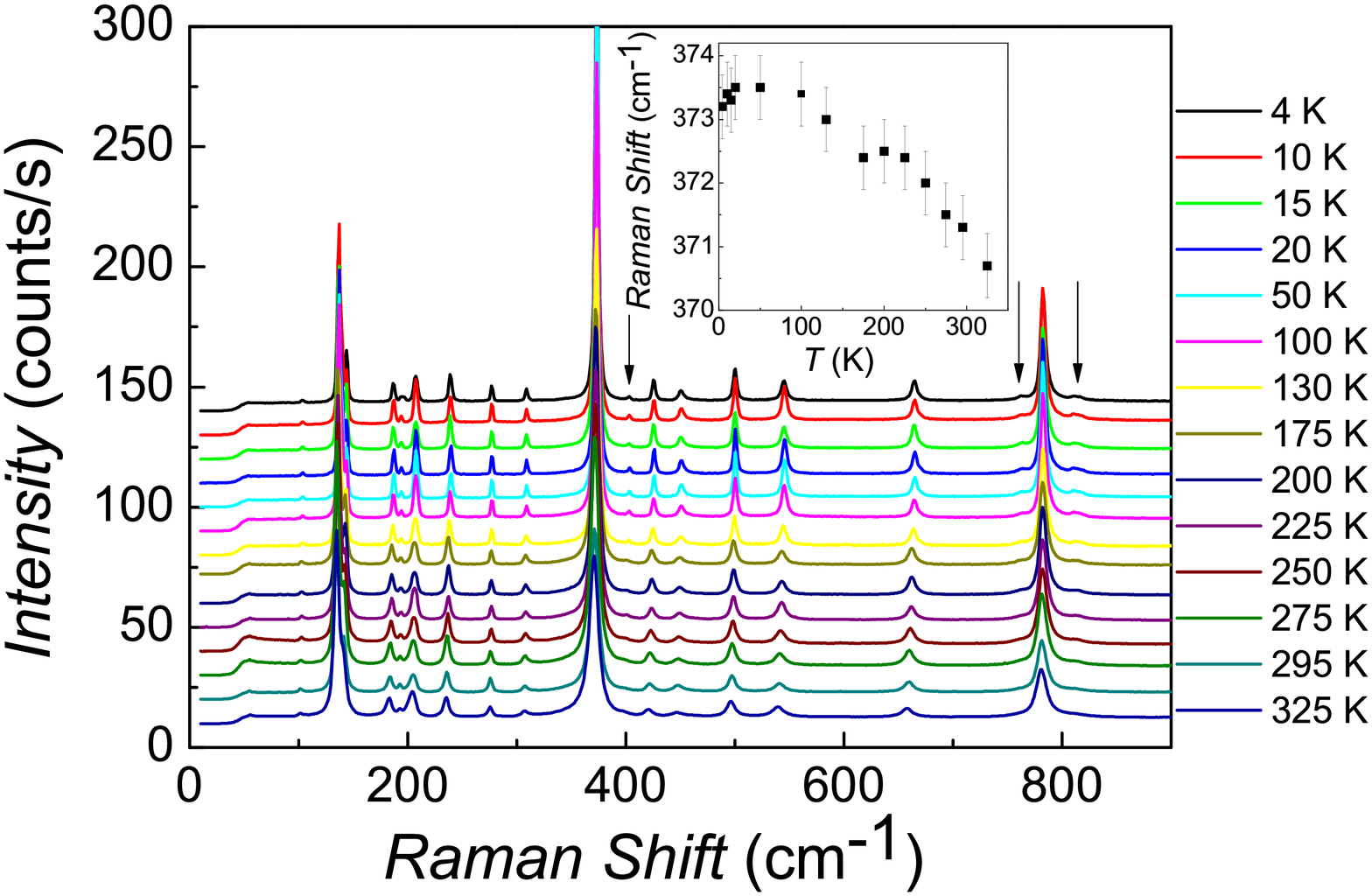}
\caption{(color online) Raman spectra of CuAs$_2$O$_4$ at various temperatures as indicated. The spectra have been shifted for clarity. The inset shows a typical down shift with increasing temperature of the 371.3 cm$^{-1}$ peak, attributed to lattice expansion.}
\label{Fig12}
\end{figure}

Raman spectroscopy is a sensitive technique which can be used  to search for structural transformations.
Raman spectra  with light polarized along the crystal $c$-axis were measured for various temperatures between 4 K and 325 K, see Figure \ref{Fig12}.
Our spectra are similar to that taken at room temperature by Kharbish\cite{Kharbish2012} with the exception of two additional peaks  detected. Table \ref{Table3} lists the results of the CuAs$_2$O$_4$ Raman spectra in comparison with data from Kharbish. The peak positions and symmetry assignments according to the GGA-LDA calculations are also given. As frequently observed for GGA and LDA calculations, the difference between the calculated and experimental wavenumbers is $\sim$ 5 - 7\%.

There are no temperature induced peak splittings indicating that the crystal symmetry, and therefore crystal structure, remains unchanged across the FM transition and down to 4 K. Overall there is a slight increase in energies as the temperature is decreased reflecting the thermal lattice contraction. Peaks positioned at $\sim$403, 763, and 812 cm$^{-1}$ (marked by arrows in Figure \ref {Fig12}) become narrower and better resolved with decreasing temperature.

\begin{table}[htp]
\caption{Comparison of the  Raman peak positions with those found by Kharbish\cite{Kharbish2012} and those obtained from GGA-LDA calculations. The notation n.o. indicates a peak which was not observed. Kharbish reported two peaks at 359 and 461 cm$^{-1}$ which were not found in the current study.}
\centering
  \begin{tabular}{ c c  c  c  c   }
\hline
\hline
    Symmetry & 295 K & Kharbish(RT)\cite{Kharbish2012} & GGA & LDA \\
    (GGA-LDA)& (cm$^{-1}$) & (cm$^{-1})$ & (cm$^{-1})$ & (cm$^{-1}$) \\
\hline
    $ B_{1g}$ & n.o. & n.o. & 23.2 & 28.9 \\
     $E_{g}$ & n.o. & n.o. & 32.5 & 35.3 \\
     $E_{g}$ & n.o. & n.o. & 98.9 & 105.3 \\
     $E_{g}$ & 101.8 & n.o. & 103.9 & 107.9 \\
     $A_{1g}$ & n.o. & n.o. & 123.6 & 128.6 \\
     $B_{2g}$ & 134.9 & 136 & 127.2 & 136.5\\
     $B_{1g}$ & 141.1 & 140 & 128.1 & 135.3 \\
     $E_{g}$ & 183.6 & 184 & 174.9 & 179.4\\
     $A_{1g}$ & 192.8 & 194 & 179.9 & 186.8 \\
     $B_{2g}$ & 204.7 & 205 & 194.5 & 204.3 \\
     $B_{1g}$ & 235.8 & 236 & 219.8 & 224.2 \\
     $E_{g}$ & 275.7 & 277 & 255.1 & 264.0 \\
     $B_{1g}$ & 307.9 & n.o. & 281.0 & 286.3 \\
     $A_{1g}$ & n.o. & n.o. & 364.3 & 370.9 \\
     $E_{g}$ & 371.3 & 372 & 374.6 & 381.4 \\
     $B_{2g}$ & 400.4 & 398 & 388.4 & 401.7 \\
     $B_{1g}$ & 421.8 & 423 & 410.7 & 419.0 \\
     $E_{g}$ & 448.1 & 448 & 439.9 & 449.1\\
     $A_{1g}$ & n.o. & n.o. & 477.4 & 481.3 \\
     $E_{g}$ & 497.3 & 498 & 496.4 & 511.8\\
     $B_{2g}$ & n.o. & n.o. & 509.9 & 514.2 \\
     $E_{g}$ & 540.8 & 541 & 535.4 & 549.0 \\
     $B_{1g}$ & 659.5 & 659 & 625.6 & 644.7\\
     $A_{1g}$ & 758.5 & 768 & 722.8 & 733.9 \\
     $B_{1g}$ & 781.5 & 782 & 731.9 & 743.2 \\
     $B_{2g}$ & 806.4 & 812 & 735.3 & 742.2 \\
\hline
\hline
  \end{tabular}
\label{Table3}
\end{table}

A contraction of the lattice will shorten the spin exchange paths, $r_i$,  between Cu spins and also alter the bonding angles, possibly differently for  $J_{\rm nn}$ and $J_{\rm nnn}$, which may lead to a small alteration of $\alpha$. However, as evidenced by the low-temperature bulk properties shown below, the ferromagnetic  spin exchange  remains dominant leading to long-range FM ordering.

\subsection{Electron Paramagnetic Resonance}
\begin{figure}[htp]
\includegraphics[width=8cm ]{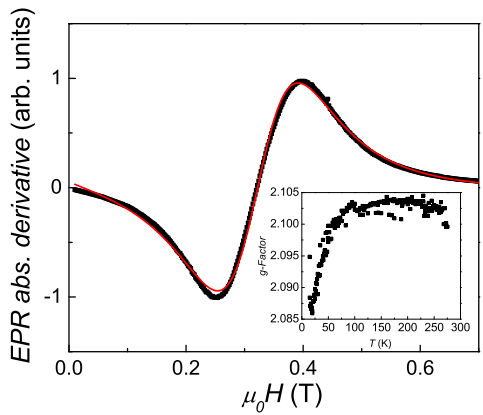}
\caption{(color online) An EPR spectrum of a polycrystalline CuAs$_2$O$_4$ sample collected at 15 K  with a microwave frequency of 9.48 GHz. The (red) solid line is a fit of the field derivative of the microwave power absorption with a single Lorentzian resonance line according to Eq. (\ref{EPRpow}). The inset displays the $g$-factor variation with temperature.}
\label{Fig13}
\end{figure}

\begin{figure}[htp]
\includegraphics[width=8cm ]{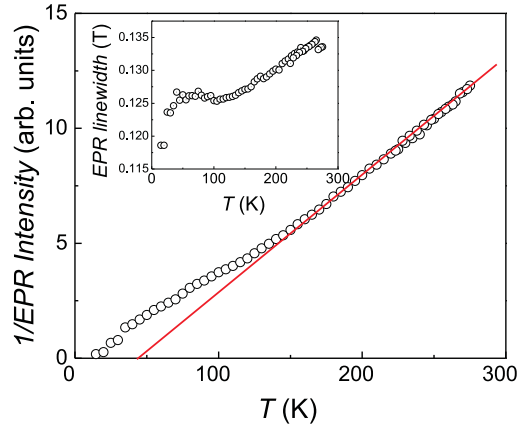}
\caption{(color online) Reciprocal   intensities of the EPR resonance lines of a polycrystalline CuAs$_2$O$_4$ sample obtained from the fits.  The (red) solid line is a linear fit of the  intensity with a Curie-Weiss like temperature dependence. The inset displays the resonance linewidth (FWHM) broadening with temperature.}
\label{Fig14}
\end{figure}

A typical EPR spectrum of a polycrystalline CuAs$_2$O$_4$ sample collected at 15 K and at a microwave frequency of 9.48 GHz is displayed in Figure \ref{Fig13}. The spectrum can be very well modeled by a field derivative of a Lorentzian absorption resonance line taking into account $\pm \omega$ resonances. The equation used to fit the spectra is as follows
\begin{equation}
\label{EPRpow}
\frac{dP_{\rm abs}}{dH} \propto \frac{d}{dH}(\frac{\Delta H + \delta(H - H_{\rm res})}{(H - H_{\rm res})^2 + \Delta H^2} + \frac{\Delta H + \delta(H+H_{\rm res})}{(H + H_{\rm res})^2 + \Delta H^2}),
\end{equation}
where $P_{\rm abs}$ is the absorbed microwave power, $\Delta H$ is the half-width at half-maximum (HWHM), $H_{\rm res}$ is the resonance field, and $\delta$ measures the degree of admixture of dispersion to the signal.
Additionally, a background offset and a linear variation of the background signal with the field were taken into account for the fits.
Very good agreement to Eq. (\ref{EPRpow}) with the data could be achieved for $\delta$ = 0, as seen in Figure \ref{Fig13}. The addition of a dispersion term ($\delta \neq$ 0) was not beneficial to the fits.
The inset in Figure \ref{Fig13} shows how the $g$-factor varies with temperature. Towards room-temperature the averaged $g$-factor was calculated to be
\begin{equation*}
 g = 2.103 \pm 0.001.
\end{equation*}
This value is close to the expected average $g$-factor for a Cu, $S$ = 1/2 system in an elongated octahedral environment and gives  an effective magnetic moment of $\mu _{\rm eff}$ = 1.82 $\mu_{\rm B}$.\cite{Abragam1970} An analysis of the inverse EPR intensity versus temperature ($T \gtrsim$ 150 K), shown in Figure \ref{Fig14}, with a Curie-Weiss type temperature of $\Theta _{\rm EPR} \sim$ 40 K. The positive $\Theta _{\rm EPR}$ indicates a predominant FM spin exchange in accordance with the positive magnetic susceptibility Curie-Weiss temperature  (see below) and the DFT calculations.
Below $\sim$ 150 K the inverse integrated  intensity bends upwards, away from the Curie-Weiss type fit. A similar behavior is also seen in the inverse susceptibility data and discussed in detail below.
The temperature dependence of the linewidth, shown in the Figure \ref{Fig14} inset, displays a broadening with temperatures above $\sim$50 K.
Below $\sim$50 K the linewidth  decreases  which we attribute to a build-up  of internal fields caused by magnetic short range ordering. These can also be a source of the temperature variation of the $g$-factor along with
minute changes of the crystal field due to the lattice contractions.


\subsection{Magnetization and Magnetic Susceptibility}
\begin{figure}[htp]
\includegraphics[width=8cm ]{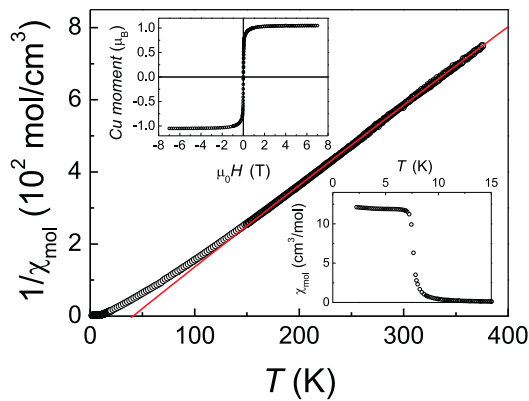}
\caption{(color online) Reciprocal magnetic susceptibility of a polycrystalline CuAs$_2$O$_4$ sample measured in a field of 1 T. The solid (red) line is a fit of the Curie-Weiss law (Eq. (\ref{eqCW})) to the data above 150 K. The lower inset shows the data below 20 K, collected at 0.01 T, in an enlarged scale. The upper inset displays the magnetization versus field collected at 1.85 K.}
\label{Fig5}
\end{figure}

The magnetic susceptibility as a function of temperature of a randomly oriented selection of CuAs$_2$O$_4$ crystals is displayed in Figure \ref{Fig5}. At high temperatures the magnetic susceptibility follows a Curie-Weiss law according to
\begin{equation}
\chi_{mol} = \frac{C}{T-\Theta_{\rm CW}}+\chi_{0}.
\label{eqCW}
\end{equation}
The Curie constant, $C$, depends on the Avogadro number $N_{\rm A}$, the spin of the system $S$ =1/2, the Boltzmann constant $k_{\rm B}$, the $g$-factor $g$,  and the Bohr magneton $\mu_{\rm B}$ according to
\begin{equation}
C = N_{\rm A}g^2{\mu_{\rm B}}^2S(S+1)/3k_{\rm B}.
\label{CWconst}
\end{equation}
The temperature independent term in Eq. (\ref{eqCW}), $\chi_0$, represents a sum of  the diamagnetic contributions, $\chi_{\rm dia}$, from the closed electron shells and the van Vleck susceptibility, $\chi_{\rm VV}$, arising from admixtures of the ground state wave functions into excited Cu electronic levels.
\begin{equation}
\chi_0 = \chi_{\rm dia} + \chi_{\rm VV}.
\label{eqchi0}
\end{equation}
From the tabulated diamagnetic increments for individual ions, $\chi_{\rm dia}$ can be estimated to contribute -77$\times$10$^{-6}$ cm$^3$/mol.\cite{Selwood1956}
The van Vleck susceptibility depends on the direction of the external field with respect to the crystal axes and the energy level separation. For a polycrystalline Cu$^{2+}$ system, it typically amounts to values between +100$\times$10$^{-6}$ cm$^3$/mol and  +120$\times$10$^{-6}$ cm$^3$/mol, resulting in a $\chi_0$ of approximately +43$\times$10$^{-6}$ cm$^3$/mol.\cite{Takigawa1989,Koo2011}

The (red) solid line in Figure \ref{Fig5} shows a fit of the experimental inverse susceptibility to Eq. (\ref{eqCW}) obtained by varying the $g$-factor and the Curie-Weiss temperature. The best fit to the data above $\sim$150 K was found with
\begin{equation*}
g = 2.15 \pm 0.01.
\end{equation*}

The Curie-Weiss temperature  from the fit was
\begin{equation*}
\Theta_{\rm CW} = 39 \pm 1 {\rm K},
\end{equation*}
indicating predominant FM spin exchange interactions, as also found by the EPR measurement.

The slight difference in the $g$-factor obtained by the magnetic susceptibility measurement to that derived from the EPR measurement (see above) may be caused by unavoidable experimental errors and/or a minor $g$-factor anisotropy .\cite{Abragam1970}

Below $\sim$8 K the magnetization rises sharply (see Figure \ref{Fig5} lower inset) indicative of a FM transition. The Curie temperature, $T_{\rm C}$, obtained from the inflection point of the susceptibility curve amounts to
\begin{equation*}
T_{\rm C} = 7.6 \pm 0.2 {\rm K}.
\end{equation*}
An isothermal magnetization measured at 1.85 K, plotted in the upper inset of
Figure \ref{Fig5}, reveals saturation of the magnetization above a field of $\sim$2.5 T with a value of 1.05$\pm$0.01 $\mu_{\rm B}$. This saturation moment is in good agreement with the expected $\sim$1 $\mu _B$ value for a $S$=1/2 system.

\begin{figure}[htp]
\includegraphics[width=8cm ]{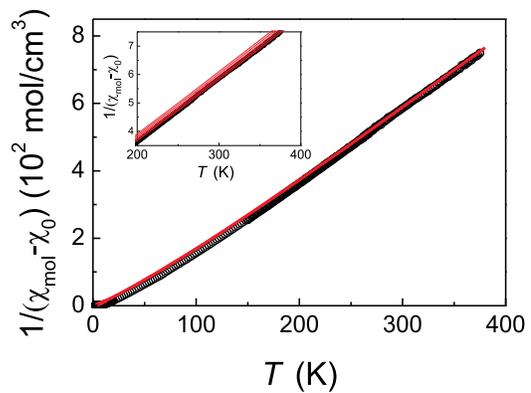}
\caption{(color online)  The inverse experimental susceptibility shown in Figure \ref{Fig5} (corrected by a temperature independent part $\chi_{\rm 0}$ = 43$\times$10$^{-6}$ cm$^3$/mol) compared with the results of the TMRG calculations, solid (red) line, for the ratio $J_{\rm nn}$/$J_{\rm nnn}$ = -4.25 (main frame). The inset displays the TMRG results, solid (red) lines, for $J_{\rm nn}$/$J_{\rm nnn}$
= -3.5, -3.75, -4, -4.25, -4.5 (from top to bottom). For all theoretical curves $J_{\rm nnn}$= -38 K and a $g$-factor of 2.155 was used. For further details see text.}
\label{Fig9a}
\end{figure}

Below $\sim$150 K the inverse susceptibility noticeably bends upwards from the high-temperature Curie-Weiss law, similar to what has also been observed in the integrated signal intensity gained from the EPR spectroscopy experiment. The temperature dependence of the susceptibility over the whole temperature range, including the upward deviation  from the high-temperature Curie-Weiss law, can be well modeled by the magnetic susceptibility of a Heisenberg chain with NN and NNN spin exchange interactions calculated with the  TMRG code as described in detail above.
Figure \ref{Fig9a} displays our experimental data in comparison with the TMRG susceptibility results calculated for ratios -4.5 $\leq J_{\rm nn}/J_{\rm nnn}$ $\leq$-3.5, a ferromagnetic NN spin exchange constant of $J_{\rm nnn}\sim$ -38 K and a $g$-factor of $g_{\rm TMRG}$=2.155, very close to the $g$-factor obtained from the Curie-Weiss fit of the high-temperature susceptibility data.

\begin{figure}[htp]
\includegraphics[width=8cm ]{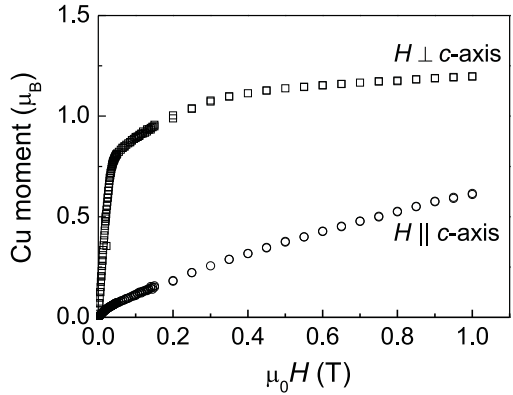}
\caption{(color online) Magnetization of an oriented CuAs$_2$O$_4$ single crystal measured at 1.85 K with the magnetic field  $H$ $\parallel$ $c$-axis and $H$ $\perp$ $c$-axis.}
\label{Fig6}
\end{figure}

Figure \ref{Fig6} displays the magnetization at 1.85 K of a single crystal
(8 $\pm$ 2 $\mu$g) oriented with the $c$-axis parallel and perpendicular to the magnetic field. With the field perpendicular to the $c$-axis, saturation  is readily achieved above $\sim$0.5 T. With the field oriented along the $c$-axis, saturation  is not obtained at 1 T indicating the $c$-axis to be a magnetic hard axis.  The saturation moment is in fair agreement with an expected value of $\sim$1 $\mu _B$, the slight excess can be attributed to errors in the mass determination of the crystal.

\subsection{Modified Arrott plots}
\begin{figure}[htp]
\includegraphics[width=8cm ]{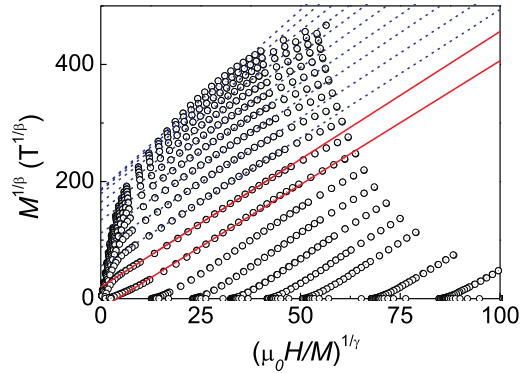}
\caption{(color online) Modified Arrott plot of the isothermal magnetization of a CuAs$_2$O$_4$ polycrystalline sample. The two solid (red) lines mark the magnetization curves measured at 7.00 K and at 7.50 K. The dashed (blue) lines mark the isotherms used to extract the zero-field magnetic polarization plot.}
\label{Fig7}
\end{figure}

An Arrott plot analysis of magnetization isotherms of a FM is a well-established method to
determine the Curie temperature, $T_{\rm C}$, and the zero-field magnetic polarization. Arrott and Noakes proposed a modified equation of state which takes into account the critical exponents, $\beta$ and $\gamma$, of the magnetization and the magnetic susceptibility, respectively. The Arrott-Noakes equation of state is given by\cite{Arrott1967}
\begin{equation}
(\mu_0 H/M)^{1/\gamma} = (T - T_{\rm C})/T_1 + (M/M_1)^{1/\beta}.
\end{equation}
$T_1$ and $M_1$ are material constants which for CuAs$_2$O$_4$ amount to $T_1$ = 0.04 $\pm$ 0.01 K and $M_1$ = 1.42 $\pm$ 0.01 T. The critical exponents can be extracted from a representation of $M^{1/\beta}$ versus ($\mu_0 H/M)^{1/\gamma}$, with the critical exponents adjusted such that the isotherms close to the Curie temperature follow a linear behavior. Such a modified Arrott-Noakes plot of CuAs$_2$O$_4$ is shown in Figure \ref{Fig7}. The critical isotherm, which extrapolates to the origin of the graph, lies between the isotherms measured at 7.00 K and 7.50 K.
The best agreement with linear behavior for the isotherms near $T_{\rm C}$ was obtained by adjusting the critical exponents to
\begin{equation*}
\beta = 0.35 \pm 0.01,
\end{equation*}
and
\begin{equation*}
\gamma = 1.32 \pm 0.01.
\end{equation*}
The critical exponent $\beta$ for the magnetization is consistent with  values for standard universality classes 3d-Heisenberg and 3d-XY, but within the experimental error does not allow for a differentiation between the two cases.\cite{Guillou1977,Guillou1980,Reisser1995}  $\gamma$ is clearly lower than the value expected for a 3d-Heisenberg model but is close to the value expected for a 3d-XY class.\cite{Guillou1977,Guillou1980,Reisser1995}
This finding is consistent  with the anisotropy seen in the single-crystal magnetization measurement  indicating an easy-plane perpendicular to the $c$-axis (see above).

\begin{figure}[htp]
\includegraphics[width=8cm ]{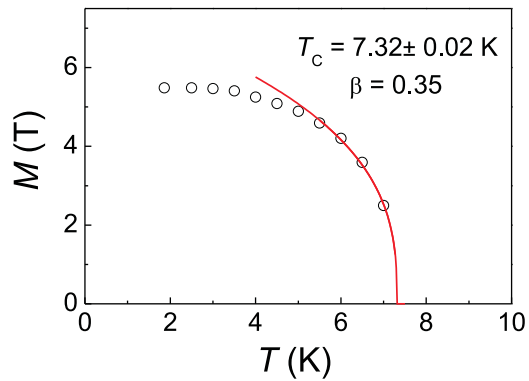}
\caption{(color online) Temperature dependence of the zero-field magnetic polarization of CuAs$_2$O$_4$. The data points were obtained from the intersections with the ordinate of the  linearly extrapolated  high-field  branches in the modified Arrott plot (i.e. $H \rightarrow$ 0).  The (red) solid line represents a critical power law fitted to the data points near $T_{\rm C}$ with a fixed critical exponent $\beta$ = 0.35.}
\label{Fig8}
\end{figure}

Figure \ref{Fig8} displays the temperature dependence of the zero-field polarization as obtained by extrapolating the high-field data in the modified Arrott plot to $H \rightarrow$ 0. By fitting a critical power law according to
\begin{equation}
M(T) = M_0 (1-T/T_{\rm C})^{\beta},
\label{eqcrit}
\end{equation}
with $\beta$ fixed to 0.35 as shown in the Modified Arrott plot, a Curie temperature of
\begin{equation*}
T_{\rm C} = 7.32 \pm 0.02 {\rm K}
\end{equation*}
is extracted. This Curie temperature is in reasonable agreement with the magnetic susceptibility data.

\subsection{ Pressure Dependence of $T_{\rm C}$}
Pressure measurements of schafarzikite, FeSb$_2$O$_4$, initially showed the cell lattice parameters to decrease linearly for pressures below 3.5 GPa. At approximately 3.5 GPa and 7 GPa, structural phase transitions occurred, inducing a symmetry reduction from spacegroups $P$4$_2$/$mbc$ via $P$2$_1$/$c$ to $P$4$_2$/$m$.\cite{Hinrichsen2006}

\begin{figure}[htp]
\includegraphics[width=8cm ]{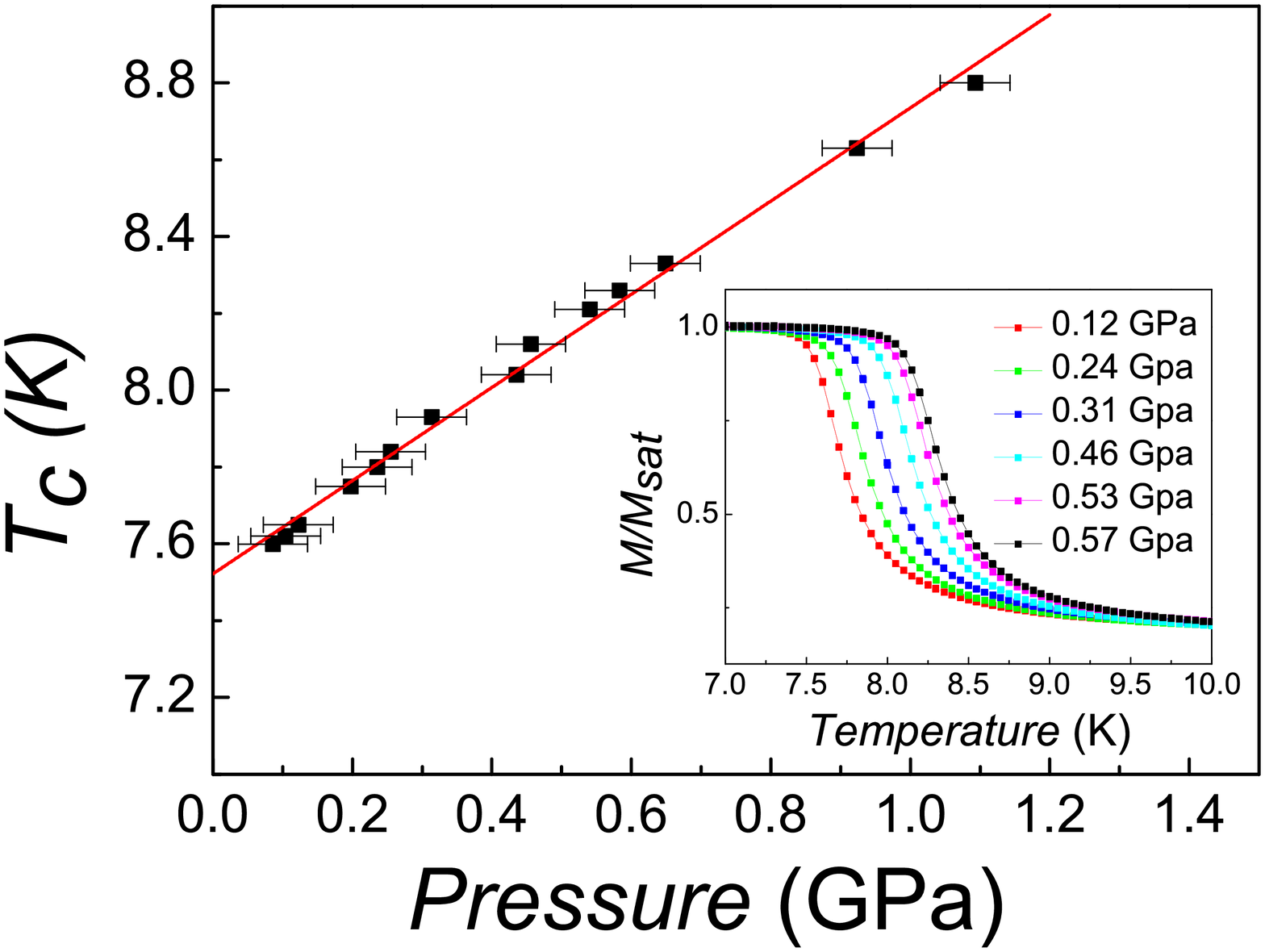}
\caption{(color online) Pressure dependence of the Curie temperature, $T_{\rm C}$, of CuAs$_2$O$_4$. The inset shows the shift of the magnetization with increasing pressures.}
\label{Fig9}
\end{figure}

Figure \ref{Fig9} displays the pressure dependence of the $T_{\rm C}$ of CuAs$_2$O$_4$ . The $T_{\rm C}$ increases linearly with pressure at a rate of
\begin{equation*}
1.35 \pm 0.05 {\rm K/GPa}.
\end{equation*}
Evidence for  pressure induced phase transitions has not been found up to 1.2 GPa.

Applying  the pressure induced decrease of the lattice parameters observed for FeSb$_2$O$_4$ similarly to CuAs$_2$O$_4$, we expect apart from the reduction of the atomic distances, a decrease of the NN Cu - O - Cu bonding angle towards 90$^{\rm o}$. The effect of pressure on the O - O - O 'buckling' angle, enclosed by the O atoms  at the edge of the basal planes running along the $c$-axis, is less pronounced.
The decrease of atomic distances and the reduction of NN bonding angles will favor the FM-NN spin exchange which explains the increase of the Curie temperature observed experimentally.\cite{Bencini1990}
Applying pressure decreases the spin exchange ratio $\alpha$, thus pushing the system further into the ferromagnetic regime.

\subsection{Heat Capacity}
\begin{figure}[htp]
\includegraphics[width=8cm ]{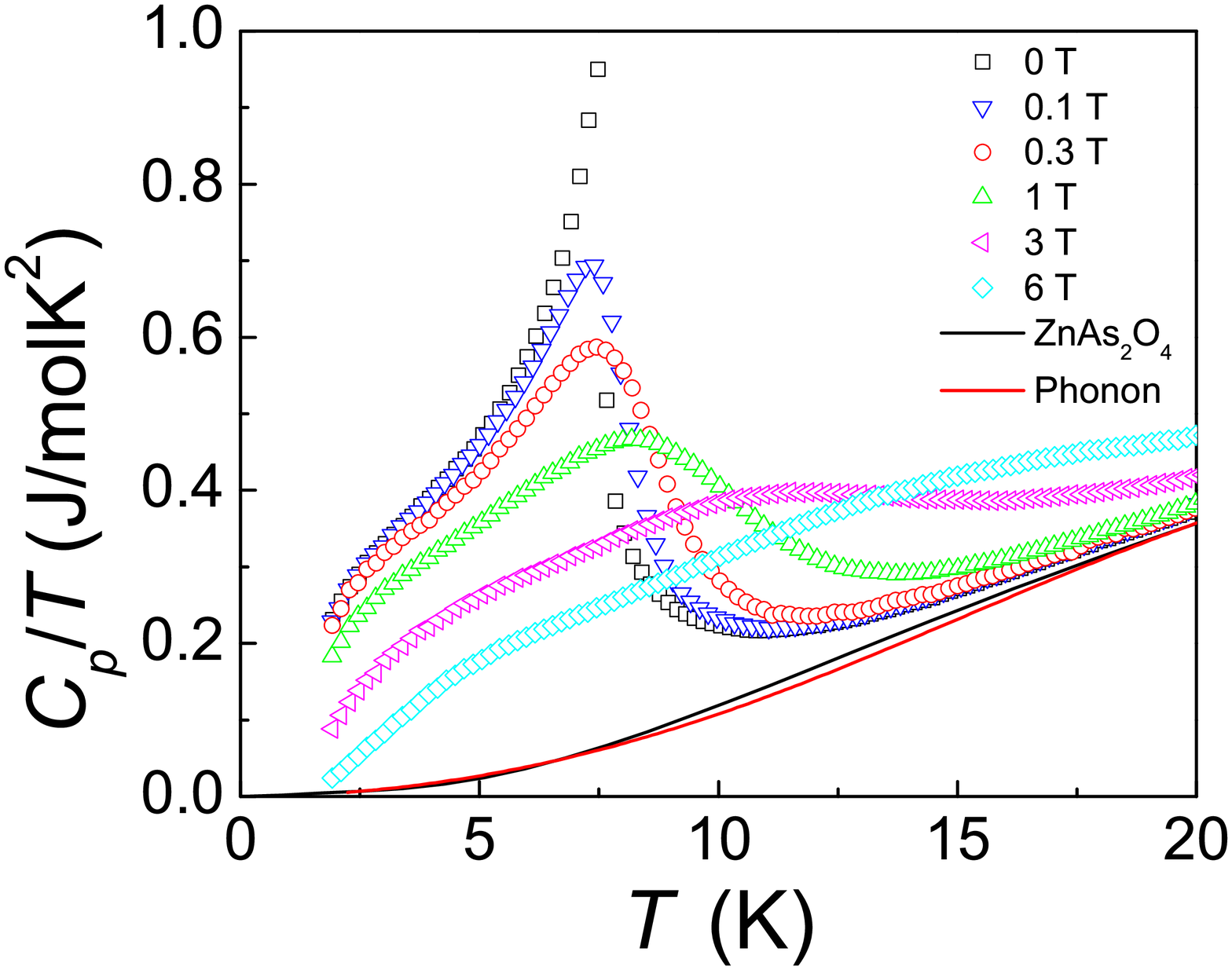}
\caption{(color online) Heat capacity of a randomly oriented ensemble of  CuAs$_2$O$_4$ crystals versus temperature and external magnetic field. The solid lines represent the scaled heat capacity of the hypothetical compound ZnAs$_2$O$_4$ and a phonon contribution to the heat capacity as obtained by extrapolating the Debye-Einstein fit to low temperatures.}
\label{Fig10}
\end{figure}

The results of the heat capacity measurements performed on a sample of randomly oriented CuAs$_2$O$_4$ crystals versus temperature and magnetic field are shown in Figure \ref{Fig10}. In zero-field, a $\lambda$-shaped anomaly is clearly exhibited at 7.4 $\pm$ 0.1K, which agrees with the preceding results from the magnetic data. Applying a magnetic field leads to a slight up-shift of the anomaly and at larger fields a broadening and suppression of the anomaly occur.
In order to subtract the lattice contribution to the heat capacity and extract the magnetic heat capacity, $C_{\rm mag}(T)$, we proceeded in two ways:
i. We approximated the lattice contribution to the heat capacity of CuAs$_2$O$_4$ by fitting  a superposition of a Debye-type and two Einstein-type heat capacity terms according to
\begin{equation}
C_P(T) = f_{\rm D} C_{\rm Deb} (\Theta_{\rm Deb},T) + \sum_i {g_i C_{{\rm Ein},i}(\Theta_{{\rm Ein},i},T)}.
\label{DebEin}
\end{equation}

The Debye-type heat capacity is given by

\begin{equation}
C_{\rm Deb}(T) = 9R (T/\Theta_{\rm Deb})^3 \int\limits_0^{\Theta_{\rm Deb}/T}\frac{x^4{\rm exp}(x)}{({\rm exp}(x)-1)^2}dx.
\label{Deb}
\end{equation}
In order to simplify the fit procedure, a Pad\'{e} approximation for the Debye-type heat capacity proposed recently by Goetsch \textit{et al.} was utilized.\cite{Goetsch2012}
The Einstein-type heat capacities, $C_{{\rm Ein},i}(T)$,  were calculated according to
\begin{equation}
C_{{\rm Ein},i}(T) = 3R (\frac{E_i}{k_{\rm B}T})^2\frac{{\rm exp}(E_i/k_{\rm B}T)}{({\rm exp}(E_i/k_{\rm B}T)-1)^2}.
\label{Ein}
\end{equation}
The  weights, $f_D$, $g_1$ and $g_2$, were conditioned such that at sufficiently high temperatures the Petit-Dulong value of 7$\times$3$R$ ($R$ is the molar gas constant) was satisfied. By fitting the weights, the Debye-temperature and two Einstein-temperatures, the experimental heat capacity above 20 K could be well approximated and extrapolated to $T  \rightarrow$ 0 K (see Figure \ref{Fig10}). The fitted parameters are summarized in Table \ref{Table2}.
\begin{table}
  \caption{Weights and characteristic temperatures used to approximate the lattice contribution to the heat capacity of CuAs$_2$O$_4$ according to Eq. (\ref{DebEin}).}
\centering
  \begin{tabular}{ c c c  }
\hline
\hline
 contribution & weight & $T$(K)\\
\hline
Debye & 1.5 & 136.59(7)\\
Einstein, $i$=1 & 2.25 & 284.5(3)\\
Einstein, $i$=2 & 3.25 & 789(1)\\
\hline
\hline
  \end{tabular}
  \label{Table2}
  \end{table}

ii. Alternatively, the heat capacity of the hypothetical compound ZnAs$_2$O$_4$ was calculated from the phonon density of states obtained by $\textit{ab initio}$ calculations (see above, Figure \ref{Fig4}) and the second derivative
of the free energy, $F(T)$. The relation is as follows
\begin{equation}
C_V(T) \approx C_P(T) =-T\left(\frac{\partial^2 F(T)}{\partial T^2}\right)_V,
\label{Eq5}
\end{equation}
where $C_V(T)$ and $C_P(T)$ are the heat capacities at constant volume and at constant pressure (accessible by the experiment), respectively.  $F(T)$ is the free energy given by
\begin{equation}
F(T)=-\int_0^\infty (\frac{\hbar\omega}{2} +  k_{\rm B}T
{\rm ln}[2n_{\rm B}(\omega)]) \rho(\omega)d\omega.
\label{Eq4}
\end{equation}
In Eq. (\ref{Eq4}), $k_{\rm B}$ represents the Boltzmann constant, $n_{\rm B}$ the
Bose-Einstein factor, and $\rho(\omega)$ the phonon density of states. The high frequency cut-off of the latter defines the
upper limit of integration in Eq. (\ref{Eq4}).

\begin{figure}[htp]
\includegraphics[width=8cm ]{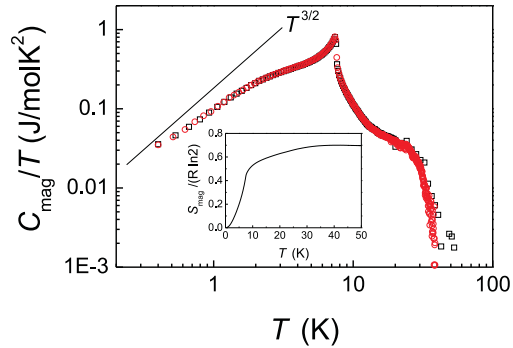}
\caption{(color online) Magnetic contribution to the specific heat of CuAs$_2$O$_4$ ($H$ = 0 T). Different symbols (red and black) indicate two independent runs of a selection of CuAs$_2$O$_4$ crystals. The inset displays the temperature dependence of the magnetic entropy obtained according to   Eq. (\ref{EqS}). The straight (black) line indicates a $T^{3/2}$ power law.}
\label{Fig11}
\end{figure}

The extracted magnetic contribution to the specific heat was obtained by subtracting the lattice contribution from the experimental results. A plot of $C_{\rm mag}(T)/T$ versus $T$ is shown in the Figure \ref{Fig11}. At low temperatures the magnetic heat capacity follows a power law
\begin{equation}
C_{\rm mag}(T) \propto T^n,
\end{equation}
with $n \sim$ 1.2, somewhat lower than expected for 3D  FM magnon ($n$ = 3/2) contributions. Above $\sim$20 K, a shoulder becomes visible which we attribute to short-range ordering contributions.

By integrating $C_{\rm mag}(T)/T$, the magnetic entropy removed by the magnetic ordering is obtained according to
\begin{equation}
S_{\rm mag}(T)=\int_0^T C_{\rm mag}(T')/T' dT'.
\label{EqS}
\end{equation}
The magnetic entropy amounts to
\begin{equation*}
S_{\rm mag} = 4.0(1) {\rm J/molK},
\end{equation*}
which is $\sim$70\% of the entropy expected for a $S$ = 1/2 system,
\begin{equation}
S_{\rm mag}=R{\rm ln}(2S+1)=R{\rm ln}(2).
\label{Smag}
\end{equation}
The largest fraction of the entropy is contained in the $\lambda$-anomaly and only a minor fraction is removed by short range ordering, the short range ordering effects were also seen in the $g$-factor temperature dependence (see above).

\section{Conclusions}\label{SectionConclus}

In summary, we have investigated the magnetic and lattice properties of CuAs$_2$O$_4$, a system characterized by axially elongated CuO$_6$ octahedra linking to form CuO$_2$ ribbon chains. \textit{Ab initio} DFT calculations  show that the nearest-neighbor intrachain spin exchange interaction is FM with a magnitude $\sim$4 times larger than the next-nearest neighbor AFM spin exchange interaction. The ratio of nearest- to  next-nearest neighbor spin exchange constants places CuAs$_2$O$_4$ in the ferromagnetic regime next to a quantum critical point between frustrated incommensurate spin-spiral and ferromagnetic order. A comparison of our temperature dependent magnetic susceptibility data with TMRG simulations supports the  \textit{ab initio} calculations.
Long-range FM ordering due to smaller interchain spin exchange interactions
is found below $\sim$7.4 K in the magnetization and heat capacity measurements.
A modified Arrott plot analysis of temperature and field dependent magnetization measurements indicates a 3d-XY critical exponents behavior. This finding is consistent with the magnetic anisotropy found by a magnetization measurement of  an oriented  single crystal. We observe that an application of external pressure increases the Curie temperature by a rate of $\sim$1.4 K/GPa, possibly due to an increase of the ferromagnetic nearest-neighbor spin exchange interaction. Raman scattering measurements show that the crystal structure is preserved across the $T_{\rm C}$ and down to 4 K. GGA and LDA calculations of the lattice dynamics have been performed and  are found to be in good agreement with the Raman spectra.

\begin{acknowledgments}
We are grateful to S. H\"ohn, E. Br\"ucher and G. Siegle for expert experimental assistance, and J. Nuss for the crystal alignment. Discussions with P. Lemmens are gratefully acknowledged. MHW thanks the NERSC Center and the HPC Center of NCSU for the computing resources.
We thank Tao  Xiang for providing the TMRG code and JML is grateful to Hantao Lu for helpful discussions.

\end{acknowledgments}

\end{document}